\documentclass[aps,prl,twocolumn,groupedaddress,showpacs]{revtex4}
\usepackage{graphicx}

\newcommand{\delete}[1]{}

\newcommand{\be}{\begin{equation}}
\newcommand{\ee}{\end{equation}}

\newcommand{\eff}{\text{eff}}

\def\beq{\begin{equation}}
\def\eeq{\end{equation}}
\def\bea{\begin{eqnarray}}
\def\eea{\end{eqnarray}}
\def\ba{\begin{array}}
\def\ea{\end{array}}

\def\ups{|\uparrow\rangle}
\def\dns{|\downarrow\rangle}
\def\aus{|{\rm{aux}}\rangle}

\begin{document}

\title{Ultra-cold mechanical resonators coupled to atoms in an optical
lattice}
\author{Andrew A. Geraci}
\email[]{aageraci@boulder.nist.gov}
\author{John Kitching}
\affiliation{Time and Frequency Division, National Institute of
Standards and Technology, Boulder, CO 80305 USA}

\date{\today}
\begin{abstract}
We propose an experiment utilizing an array of cooled
micro-cantilevers coupled to a sample of ultra-cold atoms trapped
near a micro-fabricated surface. The cantilevers allow individual
lattice site addressing for atomic state control and readout, and
potentially may be useful in optical lattice quantum computation
schemes.  Assuming resonators can be cooled to their vibrational
ground state, the implementation of a two-qubit controlled-NOT gate
with atomic internal states and the motional states of the resonator
is described.  We also consider a protocol for entangling two or
more cantilevers on the atom chip with different resonance
frequencies, using the trapped atoms as an intermediary. Although
similar experiments could be carried out with magnetic microchip
traps, the optical confinement scheme we consider may exhibit
reduced near-field magnetic noise and decoherence. Prospects for
using this novel system for tests of quantum mechanics at
macroscopic scales or quantum information processing are discussed.

\end{abstract}

\pacs{37.10.Jk,03.67.-a,07.10.Cm}

\maketitle

Over the past decade, significant research effort has been
undertaken to realize the cooling of a macroscopic mechanical
resonator to its vibrational ground state
\cite{schwab,lehnert,heidmann,tombesi,harris,bouwmeester,rugar1,Kippenberg2}.
Given the recent experimental progress in this field, ground state
cooling will likely be achieved within the next few years, and
investigating the quantum coherence in mechanical resonators
\cite{qsupmirror,qenems} will then become an exciting and important
new field of research, as the boundary between quantum microscopic
phenomena and macroscopic systems breaks down.

A natural method to study the quantum coherence in these macroscopic
systems is to observe their coupling to other quantum systems with
well understood coherence properties, for example, a two-level
system such as a Cooper-pair box \cite{schwabcpbox}, superconducting
flux qubit \cite{supercondTLSresonator}, or nitrogen-vacancy (NV)
impurity in diamond \cite{nvcenter}. Ultra-cold atoms represent
another prime example of microscopic quantum coherence, exhibiting
long coherence times. They can be used for quantum control, and can
be trapped at sub-micrometer distance from a surface \cite{vuletic}.
It has recently been pointed out that magnetic cantilevers can
couple to ultra-cold atoms at micrometer distances in a regime
analogous to the strong coupling regime of cavity quantum
electrodynamics \cite{treutlein1}, enabling studies of decoherence
and quantum control.

Trapped atoms can be arranged in regular, transportable arrays via
optical lattice potentials \cite{optlatts}. Cantilevers, on the
other hand, can be precisely defined on the surface of a chip with
lithography, and can be scaled into large two-dimensional arrays. In
this paper, we propose an experiment involving neutral atoms in
selectively occupied sites of an optical lattice near a surface,
coupled via the Zeeman interaction $H_{{\rm{int}}} = -\vec{\mu}
\cdot \vec{B}$ to a matched array of cooled magnetic
micro-cantilevers residing underneath. Unlike superconducting qubits
or NV centers, the atomic system has the feature that the atoms are
identical, and affords flexibility in that the atomic magnetic
resonances can be widely tuned with a magnetic field to match the
cantilever mechanical resonances. Also, the atoms can in principle
be transported to interact locally with multiple individual
cantilevers.

Driven magnetic resonance in a $^{87}$Rb atomic vapor has been
experimentally demonstrated with a magnetic micro-resonator
\cite{vapor}, and similar experiments are currently underway for
trapped cold-atoms \cite{treutlein1}. The demonstrated high force
sensitivity of micro-cantilevers, for example enabling single spin
detection in solids \cite{rugar2}, makes individual atom Zeeman
state detection possible at sub-micron distances. Such capabilities
may be useful for individual lattice site addressing in neutral-atom
optical-lattice quantum information processing
\cite{deutsch,bloch1,qipreview}.

\begin{figure}[!t]
\begin{center}
\includegraphics[width=0.8\columnwidth]{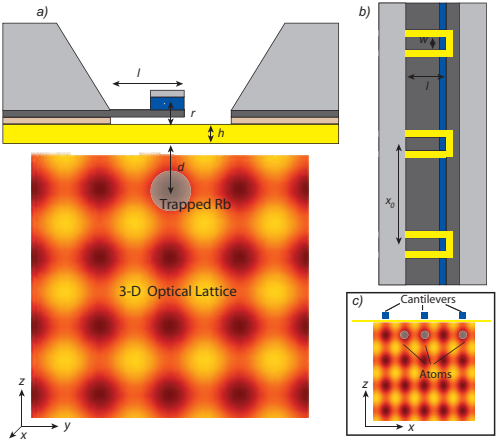}
\caption{(Color) Proposed experimental geometry. Here $r=250$ nm,
$h=150$ nm, and the total atom-cantilever vertical separation
$z=r+h+d<1$ $\mu$m. The silicon cantilevers are horizontally
separated by distance that is an integer multiple of the optical
lattice spacing $x_0=j\lambda_{\rm{eff}}/2$. Magnetic cantilever
tips and gradient compensation magnets are shown in blue.  Bulk
silicon providing structural support for the chip is shown in light
gray. Mirror membrane is shown in yellow. a),b), and c) depict views
from the $x$,$z$, and $y$ directions, respectively.} \label{chip}
\end{center}
\end{figure}
\begin{figure}[!t]
\begin{center}
\includegraphics[width=0.8\columnwidth]{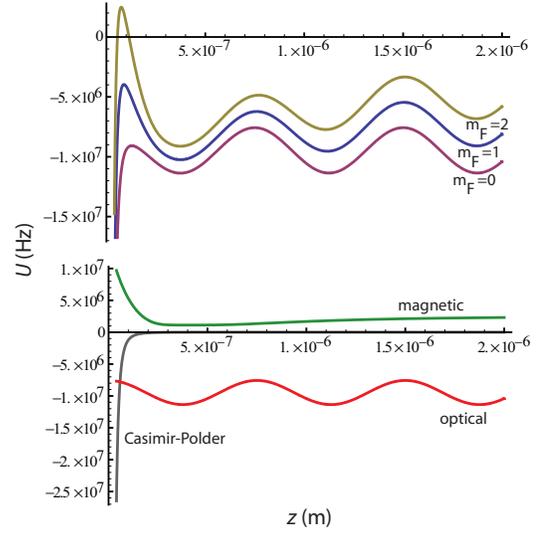}
\caption{(Color online) (Upper) Combined optical lattice,
Casimir-Polder potential, gravitational potential, and magnetic
potential of magnets and bias magnetic field as a function of
vertical distance $z$ from the surface for states $|F=2,m_F=2\rangle
\equiv \aus$, $|F=2,m_F=1\rangle \equiv \ups$, and $|F=2,m_F=0
\rangle$. (Lower) Magnetic potential for $\ups$, Casimir-polder
potential, and optical lattice shown separately. } \label{pots2}
\end{center}
\end{figure}

The proposed experimental setup is shown in Fig. {\ref{chip}}. The
cantilevers have dimensions $l=8$ $\mu$m, $w=0.2$ $\mu$m, and
$t=0.1$ $\mu$m and are separated from the 3-dimensional atomic
optical lattice by a $120$ nm thin mirror membrane coated with $30$
nm Pt. The atoms are trapped in a 2-d array at distance $d$ =
$100-400$ nm below the surface. We henceforth consider a
one-dimensional lattice of atoms and cantilevers in the $x$
direction. In principle many such arrays can be used in parallel to
form a two-dimensional lattice. The cantilevers carry a rectangular
nano-magnet with strong magnetization $M=10^6$ A/m, attainable in
thin-film magnets, and dimensions $700$ $\times$ $200$ $\times$
$150$ nm in the $x,y$ and $z$ directions, respectively. We assume a
single magnetic domain with moment in the $x$-direction determined
by the shape anisotropy. The fundamental-mode resonance frequency of
the loaded cantilever is $\omega_c / 2\pi = 1.1$ MHz. Additional
gradient compensation magnets of $x,y,z$ dimensions of $5.1$ $\mu$m
$\times$ $200$ nm $\times$ $150$ nm are located on either side of
the cantilevers with a $x$-separation of $100$ nm to minimize the
magnetic gradient at the optical potential minimum.  We take an
optical lattice with $\lambda_{{\rm{eff}}}=1500$ nm, and depth $500$
times the photon-recoil energy $E_r$, corresponding to a trap
frequency $\omega_t / 2 \pi = 124$ kHz, and lattice spacing
parameter $j=8$. For definiteness, we consider a configuration where
the first vertical anti-node occurs at $d=375$ nm from the surface.
An external bias field of $B_x = 275$ $\mu$T is applied to remove
the residual $B_x$ field for a sub-array of three cantilevers and
their neighboring compensation magnets, and $B_y=160$ $\mu$T is
applied to set the desired magnetic field and quantization axis at
the trap minimum, corresponding to a Larmor frequency $\omega_L /
2\pi$ of 1.1 MHz. The total potential is shown in Fig. \ref{pots2}.
Tunneling towards the surface can result from the decreased
potential well-depth due to the attractive Casimir-Polder
interaction \cite{casimirpolder}. The deep optical lattice serves in
part to prevent this loss mechanism, and to further avoid this loss,
we operate with only weak magnetic field seeking states, for which
the magnetic field from the cantilever tip provides a strong
repulsive interaction.

For atomic state manipulation, a local AC voltage can be applied
near a cantilever to drive its motion capacitively; the
corresponding atomic Rabi frequency is proportional to the amplitude
$\delta z$ of this motion: $ \Omega_R = \frac{g_F G_m \mu_B \delta
z}{ \hbar} $ for tip magnetic gradient $G_m$. In order to coherently
change internal atomic states, the cantilever can be driven into a
large amplitude yielding a sufficiently large Rabi frequency so that
the thermal motion has little effect over the period of a Rabi
cycle. At distance $a >> d$ from the cantilever, $G_m \propto
a^{-4}$, leading to highly localized interactions.  In addition,
differing mechanical frequencies can be used isolate neighboring
sites.

The minimum detectable force due to thermal noise at temperature $T$
is $ F_{\rm{min}} = \sqrt{\frac{4 k k_B T b}{\omega_c Q}},
\label{thn} $ where $k$ is the cantilever spring constant,
$\omega_c$ is the resonance frequency, $Q$ is its quality factor,
and $b$ is the bandwidth of the measurement. For a cantilever
separated from a single atom by a distance of 400 nm, the rms force
from atomic spin precession is $F_s = g_F \mu_B G_m / \sqrt{2} =1.9
\times 10^{-19}$ N, becoming detectable in an integration time
$b^{-1}$ of 250 ms by a thermal cantilever at $T=10$ mK, with
$Q=10^5$ and $k=0.012$
 N/m.

If we set $k_B T \approx \bar{N}_{\rm{th}} \hbar \omega_c$ and
$b=\kappa \equiv \omega_c / 2Q$, and require $F_{\rm{min}} < F_s$,
we obtain a limit on the phonon occupation number
$\bar{N}_{\rm{th}}$ of the cantilever: $\sqrt{\bar{N}_{\rm{th}}} <
\frac{\Omega_0 Q}{\sqrt{2}\omega_c}$ in order that the force be
detected within the cantilever ring-down time $2Q / \omega_c$, where
$\Omega_0 \equiv g_F \mu_B G_m z_{qm} / \hbar$.  Here $ z_{qm} =
\sqrt{\frac{\hbar}{2 m_{\rm{eff}} \omega_c}} $, and $m_{\rm{eff}}
\approx 0.24m_c + m_{\rm{mag}}$, where $m_c$ is the mass of the Si
cantilever and $m_{\rm{mag}}$ is the mass of the magnet. For
example, if $Q=10^5$, $\Omega_0= 2\pi \times 10$ rad/s, and
$\omega_c=2\pi \times 10^6$ rad/s, this requires $\bar{N}_{\rm{th}}$
of order $1$, so the cantilever must be cooled near its ground state
of motion.  To facilitate single-atom detection with a thermal
cantilever, an adiabatic fast passage protocol similar to that used
in magnetic resonance force microscopy could be used \cite{rugar1}.
A separate sub-lattice of lower frequency cantilevers could be used
for this purpose, with atoms being transportable between this
detection sub-lattice and the higher resonance frequency control
sub-lattice discussed earlier. Here the detection cantilevers can
operate at a frequency $\sim 10$ kHz corresponding to the sweep rate
of the adiabatic fast passage, while the cantilevers in the control
sub-lattice can still have resonances around $1$ MHz as discussed
earlier.  To distinguish hyperfine levels in this approach,
microwave near fields could be used, for example as proposed in Ref.
\cite{uwavepots}.

{\it{Optical lattice quantum computation.}} Single site addressing
remains a key challenge for neutral-atom optical-lattice quantum
information scenarios. Although focused laser pulses may provide a
solution \cite{focusedlasers}, the cantilever approach is not
limited by optical diffraction. As an example, we consider the SWAP
gate recently realized in Ref. \cite{xchg}. The setup consists of a
two-dimensional optical lattice that can be transformed into
double-wells. The internal qubit states used in the experiment are
the $m_F=0$ and $m_F=-1$ sublevels of the $F=1$ hyperfine ground
state manifold of $^{87}$Rb. Here a vector light-shift provides an
effective magnetic field that allows RF addressing of a sublattice
to prepare and toggle these internal states. If the optical lattice
can be formed near a surface with a lattice array of cantilevers
underneath such as that described in this work, rather than using a
global RF source for state preparation, individual driven magnetic
cantilevers could be used, adding the capability of single-site
state control.  Cantilever arrays may be generally useful for single
qubit operations in cluster-state quantum computation \cite{bloch1}.

\begin{figure}[!t]
\begin{center}
\includegraphics[width=1.0\columnwidth]{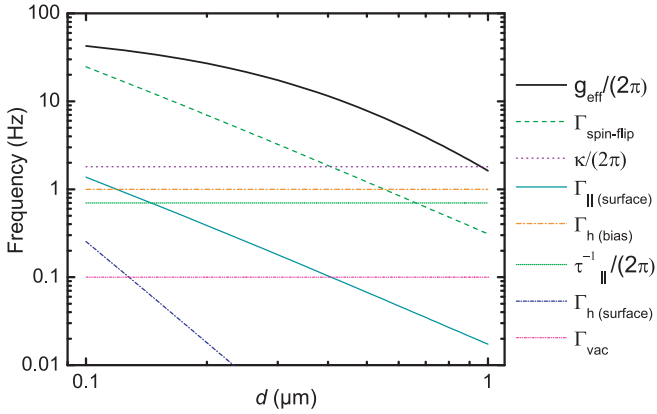}
\caption{(Color online) Rabi frequency $g_{\rm{eff}}/(2\pi)$ and
decoherence or loss rates as described in the text versus
atom-surface separation $d$. We include $\Gamma_{\rm{vac}}$ for the
estimated atomic background loss.  At $d=375$ nm the ratio of Rabi
frequency to the discussed decoherence or loss mechanisms is close
to maximal.} \label{noisefig}
\end{center}
\end{figure}

{\it{Quantum gates and entanglement.}} We now assume that the
fundamental cantilever vibrational mode can be cooled to its ground
state and that the measurement imprecision and back-action are
sufficiently small that the zero-point cantilever motion can be
detected, for example optically or capacitively from the reverse
side of the membrane. We consider the coupling of the cantilever to
a two-level system composed of the $^{87}$Rb hyperfine sub-levels
$|F=2,m_F=2\rangle$ and $|F=2,m_F=1\rangle$. The Hamiltonian
describing this coupled atom-cantilever system is \cite{treutlein1}
$ H = \hbar \omega_c (n+\frac{1}{2}) + \hbar \omega_L \hat{S_y} +
H_{\rm{int}}, $ where $ H_{{\rm{int}}} = \mu_B g_F G_m z F_z \approx
\hbar g_{\eff} (S^+ a^- + S^- a^+),$ and we have not explicitly
included the potential associated with the optical confinement or
motional degrees of freedom of the atoms. The effective single-atom
single-phonon Rabi frequency $g_{\eff} \equiv g_F G_m z_{qm} \mu_B
/\hbar$ satisfies the strong-coupling condition for the proposed
experimental parameters (see Fig. \ref{noisefig}).

The controlled-not (CNOT) gate using the coupled atom-cantilever
system is analogous to that proposed and demonstrated for trapped
ions \cite{ciraczoller,monroe}.  The qubits consist of the internal
atomic spin states $|\uparrow\rangle \equiv |F=2,m_F=1\rangle$ and
$|\downarrow\rangle \equiv |F=1,m_F=-1\rangle$, along with the
cantilever vibrational states $|0\rangle$ and $|1\rangle$
corresponding to Fock states of the ground and first excited states,
respectively.  The atomic internal state $|{\rm{aux}}\rangle \equiv
|F=2,m_F=2\rangle$ is used as an auxiliary state. By adjusting the
background magnetic field, the transition frequency $\omega_{2,1}$
between $|{\rm{aux}}\rangle$ and $|\uparrow\rangle$ can be brought
into, and out of, resonance with the fundamental mode of the
cantilever.  The non-linear Zeeman shift in a background magnetic
field of 160 $\mu$T separates $\omega_{2,1}$ from $\omega_{1,0}
\equiv |2,1\rangle \rightarrow |2,0\rangle $ by approximately $360$
Hz.

For the gate sequence, first a $\pi/2$ pulse is applied to the
spin-only component of the wave function, using a semiclassical
microwave or Raman transition with interaction Hamiltonian $
H_{\rm{hf}} = \frac{\hbar \Omega}{2} \left[\dns \langle \uparrow |
e^{-i\phi} + \ups \langle \downarrow | e^{i\phi}\right], $ for a
time $\pi/(2\Omega)$ and laser phase $\phi=\pi/2$, resulting in the
transformation $|\uparrow\rangle \longrightarrow
\frac{1}{\sqrt{2}}(|\uparrow\rangle-|\downarrow\rangle)
,|\downarrow\rangle \longrightarrow
\frac{1}{\sqrt{2}}(|\uparrow\rangle+|\downarrow\rangle).$ Then the
magnetic field is ramped so that the cantilever resonance frequency
matches $\omega_{2,1}$, and the two are allowed to interact for the
duration of a $2\pi$ pulse.  For the atom in state $\ups$, if the
cantilever is in its ground state, no interaction occurs, while if
the cantilever is in the excited state, after the $2\pi$ pulse, the
state will acquire a minus sign: $ |\uparrow\rangle|0\rangle
\longrightarrow |\uparrow\rangle|0\rangle,|\uparrow\rangle|1\rangle
\longrightarrow -|\uparrow\rangle|1\rangle. $  Finally, a $-\pi/2$
pulse is applied to the spin-only component of the wave function,
completing the CNOT operation.

The atoms in the optical lattice can act as `slow flying qubits',
and allow selective long-range entanglement between cantilevers in a
planar geometry. We assume an initial state
$|{\rm{aux}}\rangle|0\rangle |0\rangle$. The magnetic field in the
trap is brought into resonance with a cantilever at frequency
$\omega_{c1}$, and the interaction is allowed to occur for the
duration of a $\pi/2$ pulse. The resulting state is the
superposition $ \frac{1}{\sqrt{2}}(|{\rm{aux}}\rangle |0\rangle
|0\rangle - |\uparrow\rangle |1\rangle |0\rangle).$  At this point
the optical lattice is translated so the atom under consideration is
centered over a second cantilever with frequency $\omega_{c2}$. The
magnetic background field is shifted to bring this cantilever into
resonance with the Zeeman transition $\omega_{2,1}$, and they are
allowed to interact for the duration of a $\pi$ pulse. The final
state becomes $-\frac{i}{\sqrt{2}} |\uparrow\rangle [|0\rangle
|1\rangle + |1\rangle |0\rangle],$ so that the vibrational states of
the cantilevers are in an entangled superposition, despite their
differing resonant frequencies and relatively large spatial
separation.

{\it{Noise sources, Loss and Decoherence.}} We assume a cantilever
with $Q=3 \times 10^5$ at $1.1$ MHz, which has a dissipation rate
$\kappa / (2\pi)$ of $1.8$ Hz. Experimentally, $Q$ factors as high
as $3.8 \times 10^5$ have recently been achieved at low temperature
for cantilevers with resonance frequencies of order $\sim 1$ MHz
\cite{rugarhighQ}.  The cantilever decoherence time $\tau_c$
considered alone limits the atom-cantilever CNOT gate fidelity to
roughly $\sim e^{-t_g/\tau_c} \approx 0.87$ for a gate operation
time $t_g=80$ ms (for $g_{\eff}/2\pi=12.7$ Hz), and is expected to
be a dominant source of infidelity.

The atomic states involved with the controlled phase gate: $\ups$
and $\aus$, are susceptible to longitudinal magnetic de-phasing. If
$\delta \mu_{||} \approx \mu_B/2$ is the residual magnetic moment
difference between the states in the background magnetic field, the
rate of de-phasing can be estimated as $\tau_{||} \sim \hbar/ (
\delta \mu_{||} \Delta B_{||} )$ for a magnetic field fluctuation
$\Delta B_{||}$ at frequencies comparable to or less than the
inverse gate operation time \cite{sternaharonovimry}. For
longitudinal magnetic background field fluctuations at the $\sim
0.1$ nT level, the de-phasing rate $\tau_\|^{-1}$ is approximately
$0.7$ Hz. The longitudinal spectrum of thermal magnetic field noise
at low frequency is estimated \cite{varpula,chiptraplimits} as $
S_{B{||}} (\omega=0) = \frac{\mu_0^2}{32 \pi} {\frac{k_B T \sigma
h}{d(d+h)}} \label{bnoise}$ for a metal of thickness $h$ at distance
$d$ above the surface.  This leads to a de-phasing rate $\Gamma_{\|
{\rm{surface}}} \sim (\delta \mu_{||})^2 S_{B||}/2\hbar^2$, where
$S_{B||}$ is the noise spectrum of longitudinal magnetic field noise
\cite{chiptraplimits,sternaharonovimry}. The de-phasing rate for
$30$ nm of Pt with conductivity $\sigma^{-1} =10.6 \times 10^{-8}$
ohm$^{-1}$m$^{-1}$ at $T=300$ K is shown in Fig. \ref{noisefig}. Use
of a metal with higher resistivity could be beneficial. Cooling the
surface may be helpful provided the metal layer exhibits a
sub-linear dependence of resistivity on temperature.  Such
properties may be attainable by using suitable metallic alloys.

Transverse magnetic field noise at or near the atomic Larmor
frequency can result in Zeeman spin-flip transitions with a rate
\cite{henkel} $\Gamma_{if}(d,\omega_{L}) =  \Sigma_{\alpha}
\frac{\mu_B^2 g_S^2 |\langle i|\mu_{\alpha}
|f\rangle|^2}{\hbar^2}S_B^{\alpha\alpha} (d;\omega_{L}). $  Here the
initial and final states $|i\rangle$ and $|f\rangle$ are of the form
$|F,m\rangle$, $|F,m \pm 1\rangle$, and $\vec{\mu} = \mu_B g_F
\vec{F}$.  For the regime we consider, where the skin-depth at the
Larmor frequency ($\sim$ 100 $\mu$m at $1$ MHz) is much larger than
the metal thickness $h$, the spectral density of magnetic field
noise can be expressed as \cite{varpula} $ S_{B\perp}(d;\omega_{L})=
\frac{k_B T \sigma \mu_0^2}{16 \pi} \left[\frac{h}{d(d+h)}\right] =
2S_{B\|}.$   The estimated spin-flip rate $\Gamma_{\rm{spin-flip}}$
is shown in Fig. \ref{noisefig}. The rates are somewhat slower than
that ordinarily obtained for microchip traps below 1 $\mu$m since
the thickness of the metal layer is taken to be only 30 nm. This is
permissible since the metal does not need to carry the relatively
large electric currents needed for magnetic trapping on atom chips.
This feature represents a main advantage of the proposed optical
confinement scheme.

The heating rate from the trap ground state to its first excited
vibrational state by near-field noise can be estimated for a thin
metallic plane as $ \Gamma_{h \rm{(surface)}} \approx
\frac{\mu_{||}^2}{\hbar^2} \frac{z_{qm}^2}{d^2} S_{B||}(d;\omega_t)
\approx \frac{\mu_{||}^2}{\hbar^2} \frac{a_{qm}^2}{d^2} \frac{k_B T
\sigma \mu_0^2}{16 \pi} \left[\frac{h}{d(d+h)}\right],$ following
Ref. \cite{chiptraplimits}. Fluctuations in the bias fields can
result in heating at a rate \cite{chiptraplimits} $
\Gamma_{h{\rm{(bias)}}} = \frac{m_{\rm{Rb}} \omega_t^3}{2\hbar} S_h,
$ where $S_h$ is the spectrum of trap height fluctuations at the
trap frequency. For a bias field fluctuation of 0.1 nT, the
corresponding heating rate is $\approx$ 1 Hz.

We have identified a possible use for cantilevers in the context of
quantum computation with neutral atoms in an optical lattice. Also
we have described a scheme for the realization of a controlled-NOT
gate in a hybrid atomic-mechanical system. A gate fidelity
approaching $90$ $\%$ may be possible with the given experimental
parameters. Though significant experimental advances are required to
realize the quantum gate or entanglement protocols discussed in this
work, they may be possible within a few years. The system may
provide a novel testing ground for coherence in macroscopic quantum
systems. Higher gate fidelities could be achieved by employing
higher $Q$ resonators or by using tailored materials to improve the
near-field magnetic noise environment. For example, $Q$ factors in
the $10^7$ range have been observed \cite{harris} in different but
similar micro-mechanical systems.

We acknowledge helpful discussions with E. Knill, D. Wineland, J.
Home, and A. Ludlow. This work of NIST, an agency of the U.S.
government, is not subject to copyright. AG acknowledges support
from the NRC.


\begin{thebibliography}{99}

\bibitem{schwab} A. Naik {\it{et. al.}}, Nature {\bf{443}} 193 (2006).
\bibitem{lehnert} J. D. Teufel {\it{et. al.}}, Phys. Rev. Lett. {\bf{101}}, 197203 (2008).
\bibitem{heidmann} O. Arcizet {\it{et. al.}}, Nature {\bf{444}} 71 (2006).
\bibitem{tombesi} Claudiu Genes {\it{et. al.}},
Phys. Rev. {\bf{A 77}}, 033804 (2008).
\bibitem{harris} J. D. Thompson {\it{et. al.}}, Nature {\bf{452}} 72 (2006).
\bibitem{bouwmeester}
D. Kleckner and D. Bouwmeester, Nature {\bf{444}}, 75 (2006).
\bibitem{rugar1}
M. Poggio {\it{et. al.}}, Phys. Rev. Lett. {\bf{99}}, 017201 (2007).
\bibitem{Kippenberg2}
A. Schliesser {\it{et. al.}}, Nature Phys. {\bf{4}} 415 (2008).
\bibitem{qsupmirror}
W. Marshall {\it{et. al.}}, Phys. Rev. Lett. {\bf{91}} 130401
(2003).
\bibitem{qenems}
J. Eisert {\it{et. al.}}, Phys. Rev. Lett. {\bf{93}} 190402 (2004).
\bibitem{schwabcpbox}
A.D. Armour, M.P.Blencowe, and K.C.Schwab, Phys. Rev. Lett.
{\bf{88}} 148301 (2002).
\bibitem{supercondTLSresonator}
A.D. Armour and M.P. Blencowe, New J. Phys. 10, 095004 (2008), A.D.
Armour and M.P. Blencowe, New J. Phys. 10, 095005 (2008).
\bibitem{nvcenter}
P. Rabl {\it{et. al.}}, arxiv:0806.3606
\bibitem{vuletic}
Y. Lin {\it{et. al.}}, Phys. Rev. Lett. {\bf{92}} 050404 (2004).
\bibitem{treutlein1}
P. Treutlein {\it{et. al.}}, Phys. Rev. Lett. {\bf{99}}, 140403
(2007).
\bibitem{optlatts}
Y. Miroshnychenko {\it{et. al.}} Nature {\bf{442}} 151 (2006), K.D. Nelson
{\it{et. al.}}, Nature Phys. {\bf{3}} 556 (2007), P. Lee {\it{et. al}}
Phys. Rev. Lett. {\bf{99}}, 020402 (2007).
\bibitem{vapor} Y.-J. Wang {\it{et. al.}}, Phys. Rev. Lett. {\bf{97}}, 227602
(2006).
\bibitem{rugar2}
D. Rugar {\it{et. al.}}, Nature {\bf{430}}, 329-332 (2004).
\bibitem{deutsch}
G. K. Brennen {\it{et. al.}}, Phys. Rev. Lett. {\bf{82}} 1060
(1999).
\bibitem{bloch1}
O. Mandel {\it{et. al.}}, Nature 425, 937 (2003), I. Bloch, Nature
453, 1016 (2008).
\bibitem{qipreview} P. Treutlein et. al., Fortschr. Phys. 54, 702-718 (2006).
\bibitem{casimirpolder}
H.B.G. Casimir and P. Polder, Phys. Rev. {\bf{73}},360 (1948).
\bibitem{uwavepots}
P. Treutlein {\it{et. al.}}, Phys. Rev. {\bf{A 74}} 022312 (2006).
\bibitem{focusedlasers}
J. Beugnon et. al., Nat. Phys. {\bf{3}} 696 (2007).
\bibitem{xchg}
M. Anderlini, {\it{et. al.}}, Nature {\bf{448}} 452 (2007).
\bibitem{ciraczoller}
J.I.Cirac and P.Zoller, Phys. Rev. Lett. {\bf{74}} 4091 (1995).
\bibitem{monroe}
C. Monroe {\it{et. al.}}, Phys. Rev. Lett. {\bf{75}} 4714 (1995).
\bibitem{rugarhighQ}
C. Degen, private communication
\bibitem{sternaharonovimry}
A. Stern, Y. Aharonov, and Y. Imry, Phys. Rev. {\bf{A 41}} 3436
(1990).
\bibitem{chiptraplimits}
C. Henkel {\it{et. al.}}, Appl. Phys. {\bf{B}} (2002).
\bibitem{varpula}
T. Varpula, and T. Poutanen, J. Appl. Phys. {\bf{55}} 4015 (1984).
\bibitem{henkel}
C. Henkel {\it{et. al.}}, Appl. Phys. B {\bf{69}} 379 (1999).


\end{thebibliography}
\end{document}